\documentclass[prd,preprint,showpacs,amsmath,amssymb]{revtex4}
\usepackage{graphicx}
\begin{document}
\topmargin -2cm
\title{AdS--dS Stationary Rotating Black Hole
Exact Solution within Einstein--Nonlinear Electrodynamics}
\author{Alberto A. Garc\'\i a D\'\i az}
\altaffiliation{aagarcia@fis.cinvestav.mx}
\affiliation{Departamento~de~F\'{\i}sica.
\\~Centro~de~Investigaci\'on~y~de~Estudios~Avanzados~del~IPN.\\
Apdo. Postal 14-740, 07000 M\'exico DF, MEXICO.\\}

\date{\today}

\begin{abstract}
In this report the exact rotating charged black hole solution to the
Einstein--nonlinear electrodynamics theory with a cosmological
constant is presented. This black hole is equipped with mass,
rotation parameter, electric and magnetic charges, cosmological
constant $\Lambda$, and three parameters due to the nonlinear
electrodynamics: $\beta$ is associated to the potential vectors
$A_\mu$ and ${}^{\star}P_\mu$, and two constants, $F_0$ and $G_0$,
due to the presence of the invariants $F$ and $G$ in the Lagrangian
$L(F(\,x^{\mu}),G(\,x^{\mu}))$. This solution is of Petrov type D,
characterized by the Weyl tensor eigenvalue $\Psi_2$, the traceless
Ricci tensor eigenvalue $S=2\Phi_{(11)}$, and the scalar curvature
$R$; it allows for event horizons, exhibits a ring singularity and
fulfils the energy conditions. Its Maxwell limit is the de
Sitter-Anti--de Sitter--Kerr--Newman black hole solution.
\end{abstract}

\vspace{0.5cm}

\maketitle

 \tableofcontents

  \section{Introduction}
Recently the first solution for a spinning charged black hole within
Einstein--nonlinear electrodynamics theory was reported
\cite{arXivGarcia2021}. Now, its generalization with cosmological
constant of both signs is under consideration. These
 solutions are the first exact stationary
 axisymmetric solutions of the Einstein gravity
  coupled to nonlinear electrodynamics with a
 Lagrangian function $L(x^\mu)$ depending on the both invariants $F$ and
 $G$ existing in any electrodynamics, $L(F,\,G)$.
The Kerr~\cite{Kerr63}--Newman~\cite{Newman:1965my} solution is
singled out as the unique stationary axisymmetric black hole
solution of Petrov type D \cite{Petrov66} in Einstein--Maxwell
theory with $L(F)$. The Kerr and Kerr--Newman black hole solutions
are quite relevant theoretically and astrophysically. According to
the experiments to detect gravitational waves, the collision of two
massive black holes caused the emission of  the gravitational waves
registered by the LIGO experiment \cite{Abbott15} in 2015. It is
believed that nonlinear electrodynamics ought to play a relevant
role in the astrophysics of strongly magnetized objects containing
plasma as their constituent and in the behavior of magnestars.
Therefore, these new black hole solutions supporting nonlinear
electromagnetic fields could open new perspectives on the physics of
rotating celestial bodies. From the theoretical point of view these
Petrov type D solutions can be studied from various angles, but,
what is important to us, is that they may point the way to follow in
the search of stationary rotating {\bf regular} black hole
solutions. Having in mind the search of regular solutions
(gravitational structure free of singularities) we adopted the
nonlinear electrodynamics as source to the Einstein equations.
Moreover, if one looks for regular rotating charged black hole
solutions one has to proceed further to general Petov type I
metrics.

Two decades ago we reported the first exact regular spherical
symmetric back hole solution \cite {AyonGarcia:1998}, in the
framework  of Einstein--NLE equations opening a fruitful and active
area of research in the search for {\bf exact solutions}. Previous
to this work, there was known a {\it model}--the Bardeen model \cite
{Bardeen68}--whose metric determines an Einstein tensor, which, via
the Einstein equations ``{\it defines}'' a matter tensor $T_{ab}$
fulfilling the energy conditions, and therefore, one could associate
to it a viable energy matter-field tensor, although in the Bardeen
publication, no mention to a possible matter content was mentioned.
Only later \cite{AyonGarciaOnBardeen}, we succeeded to derive a
nonlinear magnetic electrodynamics source for the Bardeen model,
since then it acquires the status of exact solution in the NLE
frame.

Recently, for static spherical symmetric metrics the general exact
solution coupled to NLE \cite{Garcia-Diaz:2019acq}, with an
arbitrary structural metric function, which, via a pair of
independent Einstein equations, allows one to derive the single
associated field tensor component $\mathcal{E}$, and the
Lagrangian--Hamiltonian field function $\mathcal{L}$--$\mathcal{H}$,
which determine the entire solution, should it be singular or
regular.

There exist a geometrical approach to construct ``solutions'' of the
Einstein equations as pointed out in Stephani et al.
\cite{KramerStephani03}, pag 20: `` Any metric whatsoever is a
``solution'' of (1.1)--Einstein equations--if no restriction is
imposed on the energy--momentum tensor, since (1.1) the becomes just
a definition of $ T_{ab}$; so we must first make some assumption
about the structure of $T_{ab}$...For ``exact solution'' these
authors do not give any precise definition. Referring  to  exact
solutions we adopt the criteria of Hawking an Ellis \cite
{Hawking73}, pag 117:``We shall mean by an {\it exact solution} of
the Einstein's equations, a space--time $(\mathcal{M},g)$ in which
the field equations are satisfied with $T_{ab}$ the energy momentum
tensor of some specified form of matter which obeys postulate (a)
(`local causality') of chapter 3 and some of the energy conditions
of \S 4.3\ldots''  In this respect we consider as {\bf solutions}
those fulfilling the HE criteria of exact solution, and reserve the
name of {\bf models} for those results derived by the
``metric--defined matter tensor'' procedure, even when these models
fulfil reasonable (weak--strong) energy conditions. Therefore. in
this sense, all the reported until now Kerr--like, see
\cite{TorresFayos17} and the references therein, rotating black
holes are {\bf models}, they would become solutions, if someone
should be able to determine  the corresponding, if any, matter-field
energy--momentum tensor $T_{ab}$. In the above paragraph we are
using quotation marks although our transcriptions are partial with
minor modifications.

This exact solution describes a AdS-dS stationary rotating black
hole endowed with several parameters; it fulfils a set of four
generalized ``Maxwell equations'' for the electrodynamics fields
$F_{\mu\nu}$ and $P_{\mu\nu}$ and two independent Einstein--NLE
equations related with the two independent eigenvalues of the NLE
energy--momentum tensor. The NLE is determined by a Lagrangian
function $L(x^\alpha)$ constructed on the basis of the two
electromagnetic invariants $F(x^\alpha)$ and $G(x^\alpha)$,
$L(F,G)$, depending consequently on the coordinates $(x^\alpha)$.

\section{Review on nonlinear electrodynamics }
To avoid misinterpretations due to the use of different definitions
and sign conventions by different research groups, we give a
self--contained resume of the NLE we are dealing with. Most of this
introductory material is borrowed
from the first paragraphes of \cite{arXivGarcia2021}. \\
\noindent It is well known the standard definition of the energy
momentum tensor in general relativity: following Stephani
\cite{StephaniBook1990}, \S 9.4, from the variational principle upon
the action
\begin{eqnarray}\label{actio}
W=\int{\sqrt{-g} \,d^4x\,\left(R/2 +\kappa L_M \right)},
\end{eqnarray}
taking into account $ \delta \sqrt{-g}=\delta
g^{\mu\nu}\frac{\partial}{\partial
g_{\mu\nu}}\sqrt{-g}=\frac{1}{2}\sqrt{-g}g^{\mu\nu}\,\delta
g_{\mu\nu}$, the variation yields
\begin{eqnarray}
\delta W=-\frac{1}{2}\int{(R^{\mu\nu}-\frac{1}{2}R\,g^{\mu\nu}
-\kappa\,T^{\mu\nu})\,\sqrt{-g} \delta g_{\mu\nu}}\,d^4x.
\end{eqnarray}
Hence from the vanishing of this variation one arrives at the
Einstein equations in the presence of a cosmological constant,
coupled to  matter and fields described by an energy--momentum
tensor $T^{\mu\nu}$
\begin{equation}\label{Einstein}
R^{\mu\nu}-\frac{1}{2}R\,g^{\mu\nu}-\Lambda g^{\mu\nu}=\kappa
T^{\mu\nu};\,
T^{\mu\nu}=\,\frac{2}{\sqrt{-g}}\frac{\delta(\sqrt{-g}\,L_M)}{\delta\,g_{\mu\nu}},
\end{equation}
where $L_M$ stands for the matter--field Lagrangian $L_M$; the
cosmological constant arises from the contribution in the action of
the $\Lambda$--term, $L_{\Lambda}=-{\Lambda}/{\kappa}$,
$\,\kappa\,\,T^{\mu\nu}=-\Lambda g^{\mu\nu}$.\\

The electrodynamics is described by an antisymmetric electromagnetic
field tensor $F_{\mu \nu}=A_{\nu ,\mu}-A_{\mu ,\nu}$, together with
its dual field tensor
\begin{eqnarray}
\label{dualF}
{}^{\star}F_{\alpha\beta}=\frac{1}{2}\eta_{\alpha\beta\mu\nu}F^{\mu\nu}
=\frac{1}{2}\sqrt{-g}\epsilon_{\alpha\beta\mu\nu}F^{\mu\nu},\nonumber\\
{}^{\star}F^{\alpha\beta}=\frac{1}{2}\eta^{\alpha\beta\mu\nu}F_{\mu\nu}
=-\frac{1}{2\sqrt{-g}}\epsilon^{\alpha\beta\mu\nu}F_{\mu\nu}
\end{eqnarray}
where the numerical $\epsilon$--tensor is associated to the
4--Kronecker tensor.\\
 These tensors, being antisymmetric, allow for two invariants
  $F$ and $G$ (pseudo--scalar):
\begin{equation}\label{InvFG}
F=\frac
{1}{4}F_{\mu\nu}F^{\mu\nu},\,{{G}}=\frac{1}{4}{}^\star{{F}}_{\mu\nu}F^{\mu\nu},
{}^\star{{F}}_{\mu\sigma}F^{\nu\sigma} ={{G}}\,\delta^\nu_\mu.
\end{equation}
In nonlinear electrodynamics the Lagrangian function $L$ depends on
these electromagnetic invariants $F$ and $G$ , $L=L(F,\,G)$. To
construct the energy--momentum tensor $T^{\mu\nu}$ one accomplishes
the variation of $L_M=-L(F,G)$  with respect to $g_{\mu\nu}$ of the
Lagrangian function, which yields
\begin{eqnarray}\label{EnegryT1}
T^{\mu\nu}&&=-L\, g^{\mu\nu}+L_F\,F^{\mu\sigma}
{F^{\nu}}_{\sigma}+L_G{F^{\mu\sigma}}{}^\star{F^{\nu}}_{\sigma}\nonumber\\&&=:-L\,
g^{\mu\nu}+F^{\mu\sigma}{P^\nu}_{\sigma},
\end{eqnarray}
where $P_{\mu\nu}$ is a new electromagnetic field, which always can
be introduced in nonlinear electrodynamics, namely
\begin{eqnarray} \label{Pfield}
P_{\mu\nu}=2\frac{\partial L}{\partial F^{\mu\nu}}=L_F
F_{\mu\nu}+L_{{G}} {}^{\star}{F}_{\mu\nu},
\end{eqnarray}
which one identifies as the $P_{\mu\nu}$ field tensor; see
$p^{kl}$--field tensor of Born--Infeld \cite{BornInfeld1934}
Pleba\'nski \cite{PlebLect}, see \cite {SalazarGarciaPleb1} too.

From the antisymmetric tensor field tensor $P_{\mu\nu}$ one
constructs its dual field tensor ${}^\star P_{\mu\nu}$ and the
invariants
\begin{eqnarray}\label{dualPInvPQ}
&&{}^{\star}P_{\mu \nu}:=\frac{1}{2}
\sqrt{-g}\epsilon_{\mu\nu\alpha\beta}P^{\alpha
\beta},\, \,{{}^\star{P^{\alpha\beta}}}=-\frac{1}{2\sqrt{-g}}
\epsilon^{\alpha\beta\mu\nu}P_{\mu\nu}, \nonumber\\&&
P=\frac{1}{4}P_{\mu\nu}P^{\mu\nu},\,{{Q}}=\frac{1}{4}
{{}^{\star}{P}}_{\mu\nu}P^{\mu\nu}.
 \end{eqnarray}
The energy--momentum tensor (\ref{EnegryT1}) allows for a similar
writing in term
 of the Hamiltonian function $H(P,Q)$,
 ${}^{\star}F_{\mu\nu}$, and  ${}^{\star}P_{\mu\nu}$  fields.
 Replacing in the definition of the energy--momentum tensor
 (\ref{EnegryT1}) the relation
 \begin{eqnarray}\label{Potrel}&&
 {}^{\star} P_{\mu\sigma}{}^{\star}
 F^{\sigma\nu}=\frac{1}{4}{\epsilon}_{\sigma\mu\alpha\beta}
 {\epsilon}^{\sigma\nu\lambda\rho}  P^{\alpha \beta}F_{\lambda\rho}
 =-\frac{3!}{4} {\delta^\nu}_{[\mu }
 {\delta^\lambda}_\alpha {\delta^\rho}_{\beta]} P^{\alpha
 \beta}F_{\lambda\rho}\nonumber\\&&
 =\frac{1}{4} \left(4\,F_{\lambda\mu} P^{\lambda\nu}-2{\delta^\nu}_{\mu}
 P_{\alpha\beta}  F^{\alpha\beta} \right)=  \,F_{\lambda\mu}
 P^{\lambda\nu}-\frac{1}{2} {\delta^\nu}_{\mu}
 P_{\alpha\beta}  F^{\alpha\beta} ,
  \end{eqnarray}
one arrives at
      \begin{eqnarray}\label{enerTen}&&
  T_{\mu\nu}=-\left(L-\frac{1}{2}\,F_{\alpha\beta}
  F^{{\alpha\beta}} \right)\, g_{\mu\nu} - {}^{\star}P_{\mu\alpha}
  {{}^{\star}F_\nu}^{\alpha}=:H-{}^{\star}P_{\mu\alpha}
  {{}^{\star}F_\nu}^{\alpha} .
 \end{eqnarray}
Thus one arrives at the ``Hamiltonian function'' $ H(P,Q)$, in the
terminology of Born--Infeld--Pleba\'nski,  associated with the
Lagrangian function $L(F,G)$ via
\begin{eqnarray}\label{Legandre}
L(F,G)=\frac{1}{2}F_{\mu\nu}P^{\mu\nu}-H(P,Q),
\end{eqnarray}
which is known as a Legandre transformation, with:
\begin{eqnarray}\label{potentialdef} &&
P_{\mu\nu}=2\frac{\partial L}{\partial F^{\mu\nu}}=L_F\,
F_{\mu\nu}+L_G \,{{}^\star F}_{\mu\nu},\nonumber\\&&
F_{\mu\nu}=2\frac{\partial H}{\partial P^{\mu\nu}}=H_P\,
P_{\mu\nu}+H_{{Q}} {}^{\star}{P}_{\mu\nu}.
\end{eqnarray}
The Legandre transformation (\ref{Legandre}) is a consequence of the
relation (\ref{Potrel}) to determine the energy--momentum
tensor (\ref{enerTen}) in term of the Hamiltonian function $H(P,Q)$.\\
The electrodynamics is determined through  the ``Faraday--Maxwell''
electromagnetic field equations, which in vacuum are
\begin{eqnarray}\label{Faraday}
{{}^\star{F^{\mu\nu}}}_{;\nu}=0 \rightarrow {(\sqrt{-g}{}^\star {F^{\mu\nu}})_{,\nu}=0 },
\end{eqnarray}
\begin{eqnarray}\label{Maxwell}
{P^{\mu\nu}}_{;\nu}=0
\rightarrow{\left[{\sqrt{-g}L_F\,{F^{\mu\nu}}+\sqrt{-g}L_G{{}^\star
{F^{\mu\nu}}}}\right]_{,\nu}=0},
\end{eqnarray}
which can be written by means of a closed 2--form $d\omega=0$,
\begin{eqnarray*}
\omega=\frac{1}{2}\left(F_{\mu\nu}+{{}^\star
P_{\mu\nu}}\right)dx^\mu\,\wedge
dx^\nu=\frac{1}{2}\left(F_{ab}+{{}^\star P_{ab}}\right)e^a
\wedge\,e^b,\
\end{eqnarray*}
since $F_{\mu\nu}$ and ${{}^\star P_{\mu\nu}}$ are curls.

Moreover, the Lagrangian function is assumed to be an integrable
function of the coordinates $x^\alpha$ of the 1--form equation
 \begin{eqnarray}&&\label{lagran}
d\,L(F(x^{\alpha}),G(x^{\alpha}))=\frac{\partial L}{\partial
F}\frac{\partial F}{\partial x^{\alpha}} d\,{x^\alpha}+
\frac{\partial L}{\partial G}\frac{\partial G}{\partial x^\alpha}
d\,{x^\alpha} =\frac{\partial L}{\partial x^\alpha}\, d\,{x^\alpha}
.
\end{eqnarray}

   \subsection{ Coordinate--dependent nonlinear electrodynamics}

In the nonlinear electrodynamics under consideration is assumed the
existence of two pair of electromagnetic antisymmetric fields
$F_{\mu\nu}$ and its dual ${}^{\star}F_{\mu\nu}$ and a nonlinear
associated (to be established) second pair of antisymmetric field
$P_{\mu\nu}$ and its dual ${}^{\star}P_{\mu\nu}$. Each pair allow
for two sets of eigenvalues, such that each set is constituted by
two pair of different eigenvalues, hence. in the corresponding
eigenvector basis the field tensors are symbolically represented as
$(F_{\mu\nu})=\text{diag}(\lambda_1,\lambda_1,\lambda_2,\lambda_2)$
and $(P_{\mu\nu})=\text{diag}(\pi_1,\pi_1,\pi_2,\pi_2)$.
Consequently, there arise the invariants $F$ and $G$ for the field
$F_{\mu\nu}$ and
 $P$ and $G$ a for the field $P_{\mu\nu}$, see definitions above,
 with these secondary objects one may build the
 Lagrangian$ L(F(x^\alpha),G(x^\alpha) )$ or the Hamiltonian $ L(P(x^\alpha),Q(x^\alpha)
 )$ formulations.\\
In what follows we focus on the Lagrangian formulation of the
nonlinear electrodynamics by fixing the relations between the fields
through (\ref{potentialdef}), $P_{\mu\nu}=2\frac{\partial L
}{\partial F^{\mu\nu}}$
 and $F_{\mu\nu}=2\frac{\partial L }{\partial P^{\mu\nu}}$.
At this level the structure is quite general; the constraints arise
by requiring the fulfillment of the ``Maxwell'' equations
(\ref{Maxwell})
 \begin{eqnarray*}\label{Maxwellx}
{P^{\mu\nu}}_{;\nu}=0
\rightarrow{\left[{\sqrt{-g}L_F\,{F^{\mu\nu}}+\sqrt{-g}L_G{{}^\star
{F^{\mu\nu}}}}\right]_{,\nu}=0},
\end{eqnarray*}
stating that ${{}^{\star}P_{\mu\nu}}$ is a
 curl,
 ${}^{\star}P_{\mu\nu}={}^{\star}P_{\nu,\mu}-{}^{\star}P_{\mu,\nu}$; the $F_{\mu\nu}$ is a curl
 $F_{\mu\nu}=A_{\nu,\mu}-A_{\mu,\nu}$ and therefore its dual fulfils  ${{}^{\star}F^{\mu\nu}}_{;\nu}
 =0$. The
 above equations allow the determination of the functions
 $L_F(x^\alpha)$ and $L_G(x^\alpha)$, depending on the coordinates, in terms of the field
 components of $F_{\mu\nu}$, $ {}^{\star}F_{\mu\nu}$ ,$P_{\mu\nu}$ , and ${}^{\star}P_{\mu\nu}
 $. The determination of the Lagrangian $L(x^\alpha)$ in terms of the
 coordinates is achieved by the fulfillment of the closure condition  $d^2\,L=0$ of the 1--form equation
 \begin{eqnarray*}&&\label{lagranX}
d\,L(F(x^{\alpha}),G(x^{\alpha}))=\frac{\partial L}{\partial
F}\frac{\partial F}{\partial x^{\alpha}} d\,{x^\alpha}+
\frac{\partial L}{\partial G}\frac{\partial G}{\partial x^\alpha}
d\,{x^\alpha} =\frac{\partial L}{\partial x^\alpha}\, d\,{x^\alpha}
.
\end{eqnarray*}
The energy--momentum tensor for the nonlinear electrodynamics under
consideration is derived via the standard definition
\begin{equation*}\label{EinsteinX}
T^{\mu\nu}=-\,\frac{2}{\sqrt{-g}}\frac{\delta(\sqrt{-g}
\,L(x^\beta))}{\delta\,g_{\mu\nu}}=-L(x^\beta)\, g^{\mu\nu}
+{F^\mu}_{\alpha} {P}^{\nu\alpha},
\end{equation*}
with the eigenvalue structure $ { T^{a}}_b=\text
{diag}(\tau_1,\tau_1,\tau_2,\tau_2) $, which is the eigenvalue
characteristic of any electrodynamics.\\

 There exit some researchers
in electrodynamics in relativity that sustain the point of view that
if the Lagrangian of the theory is not constructed in terms of the
invariants $F$ and $G$ alone the theory is spurious, contrary to the
view point that I expose in details above, in which the fundamental
cornerstones are the electromagnetic fields $F^{\mu\nu}$ and
$P^{\mu\nu}$ and their dual ones, which, being solutions of the
field equations are coordinate--dependent quantities; the secondary
structural functions, i.e., the invariants $F$ and $P$ occur to
functions of the coordinates, finally, the Lagrangian $L$ integrates
its 1--form equation as a function of coordinates $x^{\alpha}$. This
wider point of view, although is contained in the
 Born--Infeld--Pleba\'nski formulation via the simple recognition--acceptance of
 the coordinate description, can be called ``Coordinate Dependent Non
 Linear Electrodynamics.''

  This resume explains the constructive way of
 searching for solutions in nonlinear electrodynamics to be
 used in this work
 in the integration of the Einstein equations; one first determines
 the electromagnetic fields and latter one constructs the invariants and
 the Lagrangian of the theory to proceed further with the Einstein
 equations. In this approach the problem of representing the
 Lagrangian  purely in terms of the invariants could become
 insolvable, because of the appearance of possible transcendent relations of
 invariants versus coordinates, or polynomials of degree
higher than five, Abel--Galois restrictions, among others.\\

Only for some simple cases of static spherically symmetric solutions
of Einstein--NLE one can construct a posteriori the explicit
dependence $L(F)$. In the stationary axisymmetric case, the
Kerr--Newman is the only known until now case where the relation
$L(F)$ can be established, see details in this text. The new NLE
solution presented here gives rise to transcendent relations of
eight degree and consequently insolvable to establish of the
dependence of the coordinates as functions of the invariants, thus a
stationary axisymmetric generalization of the Lagrangian $L=L(F,G)$
for a NLE generalization of the Kerr--Newman solution is quite
difficult to guess. On this respect one can recall the guessing
process followed to ``derive'' the Newman et al. solution
\cite{Newman:1965my}, which consists in the application of
Janis--Newman procedure, see comments on this respect in Pleba\'nski
et al.\cite{KrasinkiPle06}, pag 458 , ``Historically, the
generalization (of the Kerr solution) for an electric charge was
discovered by Newman et al.(1965) by a procedure equally mysterious
as the derivation of the Kerr metric itself....'' In the Stephani
book \cite { Stephani82}, pag 230, one reads: ``Since its
mathematical structure is rather complicated, we shall not construct
a derivation from the Einstein field equations.'', of the Kerr
solution. As a by product, we derive here, from results of the
coordinate dependent method in NLE, the Kerr--Newman solution
determining first the electromagnetic vector potential, next, the
Lagrangian function $L(x^\mu)$
 and ,finally, establishing the relation $L(F,G)$. Incidentally, in our
 approach for searching stationary
axisymmetric solutions, in the linear Maxwell case $L=F$, the vector
potential fulfils the Laplace equation.

The integration process for a given Lagrangian $L=L(F,G)$ in terms
of $F$ and $G$ requires the integration of the ``conservation''
field equations $ {{}^{\star}F^{\mu\nu}}_{;\nu}=0$ and  $
{P^{\mu\nu}}_{;\nu}=0$, once this  is done, one replaces the
electromagnetic fields through the invariants (using the derivatives
$L_F$ and $L_G$) into the recently integrated field equations to
proceed further with the integration of the vector potentials
$A_\mu$ and ${}^{\star}P_\mu$, which, in many cases, cannot be
integrated at all.

 \subsection{ Summary on nonlinear electrodynamics in relativity}

 Although it may sound repetitive, I would insist and emphasize on the eigenvalue structure
 of the electrodynamics, no matter the names that one coins for its
 different variants, to differentiate it from the ``anisotropic''
 fluid structure.
 By extension of the Maxwell field theory from special to general relativity,
 the electromagnetic field
 tensor $F_{\alpha \beta}$ is assumed to be antisymmetric
 $F_{\alpha \beta}=A_{\beta, \alpha}-A_{\alpha, \beta}$ depending on a vector
 potential $A_{\alpha}$, thus its tensor matrix allows for two pair
 of different eigenvalues, symbolically
 $({F^\alpha
 }_{\beta})=\text{diag}(\lambda_1,\lambda_1,\lambda_2,\lambda_2)$,
 associated to $F_{\alpha \beta}$  and ${}^{\star}F_{\alpha \beta}$,
 (\ref{dualF}).
 These eigenvalues are related with the field invariants $F$ and
 $G$, (\ref{InvFG}). In the field formulation of the Maxwell theory, the Lagrangian
 function $L$ is assumed to depend on $F$. In its nonlinear
 electrodynamics
 generalization (Born--Infeld--Pleba\'nki ) the Lagrangian $L$
 depends on both invariants $F$ and $G$. There appear in the NLE
 theory two new field tensors $P_{\alpha \beta}$ (\ref{Pfield})  and its dual ${}^{\star}P_{\alpha
 \beta}$ (\ref{dualPInvPQ}), with their own pair of eigenvalues $\Lambda_1$ and
 $\Lambda_2$, such that  $({P^\alpha
 }_{\beta})=\text{diag}(\Lambda_1,\Lambda_1,\Lambda_2,\Lambda_2)$,
  and correspondingly two field invariants $P$ and $Q$,(\ref{dualPInvPQ}).
 The energy--momentum tensor $T_{\mu\nu}$ occurs to be given by
 (\ref{EnegryT1}), and as consequence of its structure, its
 eigenvalue problem solves for two pairs of different eigenvalues such that $({T^\alpha
 }_{\beta})=\text{diag}(\tau_1,\tau_1,\tau_2,\tau_2)$, and
 consequently belongs to the Segre((11)(1,1))-Pleba\'nski
 $(2S-2T)_{(11)}$ class~\cite{PlebClass},
  the family of electromagnetic energy--momentum tensors.
 Thus, the nonlinear electrodynamics can be defined by the invariant
 eigenvalue properties of its constitutive fields obeying the
 corresponding field equations (\ref{Faraday}) and (\ref{Maxwell}),
 and the corresponding energy--momentum tensor. \\
  On the other hand, taking into account the relation
${F}^{\mu\sigma}{{}^\star F}_{\nu\sigma}=G\delta^\mu_\nu$, one
determines the traceless NLE energy--momentum tensor
$\Upsilon_{\mu\nu}$  to be
\begin{eqnarray}&&\label{Upsilon}
{\Upsilon^\mu}_{\nu}:={T^\mu}_{\nu}-\frac{T}{4}{\delta^\mu}_{\nu}=
L_F({F}^{\mu\sigma}{F}_{\nu\sigma}-\,F{\delta^\mu}_{\nu}),\,\,T:={T^\mu}_{\mu}=-4\,L
+4\,L_F\,{F}+4\,L_G\,G.
\end{eqnarray}
Of course, this traceless energy tensor, via the Einstein equations
is equivalent to the traceless Ricci tensor
${S^\mu}_{\nu}=\kappa{\Upsilon^\mu}_{\nu}$. Consequently it falls
into the Segre((11)(1,1))-Pleba\'nski $(2S-2T)_{(11)}$ class of
energy tensors, see Stephani et al. \cite{KramerStephani03}, Chapter
5. In the linear Maxwell limit, $ ({T^a}_b)= {\text
{diag}}(\lambda_1,\lambda_1,-\lambda_1,-\lambda_1)
 $, see  \cite{LichBook1955},pag 20.\\

\subsection{Definition of exact solution to nonlinear
electrodynamics equations in Einstein--NLE theory}

 By this definition an exact solution means that the tensor field
$F_{\alpha\beta}$, and $P_{\alpha\beta}$, and
 their dual ${}^{\star}F_{\alpha\beta}$, and ${}^{\star}P_{\alpha\beta}$
fulfil the set of the coordinate `` Faraday--Maxwell''
equations,(\ref{Faraday}) and (\ref{Maxwell}), and the Lagrangian
1--form (\ref{lagran}) fits the closure condition $d^2L=0$. An exact
solution to Einstein--NLE theory is a solution that fulfil, besides
the electrodynamics equations, the Einstein equations
(\ref{Einstein}) equated to the nonlinear electrodynamics
energy--momentum tensor (\ref{EnegryT1}).

This definition is wider than the one requiring for the weak Maxwell
limit $L\simeq F+\mathcal{O}(F^2,G^2)$. The Lagrangian function
derivation in terms of coordinates we considered more efficient and
constructive than the description at initio  through the two
invariants $F$ and $G$, this because the field tensor description is
done, at the end at day, in the language of coordinates.

 \section{Metric and tetrads}
We present the first stationary axially symmetric exact black hole
solution to the Einstein equations coupled to nonlinear
electrodynamics in the presence of a (positive or negative)
cosmological constant given by the metric \marginpar{DONE}
\begin{eqnarray}&&\label{metric1}
ds^2=\,\frac{a^2\sin^2{\theta}}{\rho
^2}\,({1-\frac{\Lambda}{3}a^2\,\cos^2{\theta}})\,\left( {\bf{dt}}
-\frac{a^2+r^2}{a} {\bf d \phi} \right)^2\nonumber\\&&+\frac{\rho
^2}{Q(r)}\,{\bf dr}^2+\frac{\rho
^2}{1-\frac{\Lambda}{3}a^2\,\cos^2{\theta}}\,{\bf d \theta}^2
\nonumber\\&& -{\frac{Q(r)}{\rho ^2}}\left( {\bf{dt}} -
a\sin^2{\theta}{\bf d \phi} \right)^2,\,
\rho(\theta,r):=\sqrt{{r^2+a^2\cos^2{\theta}}}.\nonumber\\&&
\end{eqnarray}
The structural function $Q(r)$ is the single metric function to be
determined by solving the Einstein equations, occasionally we use
its representation through the auxiliary function $K(r)$
\begin{eqnarray}\label{solQbeta}&&
Q(r) =K(r)-2\,m\,r+{r}^{2}+{a}^{2}+\frac{\Lambda}{3}\,r^2(r^2+a^2).
\end{eqnarray}
The null tetrad $\bf{e^a}$ used is
\begin{displaymath}\label{tetradn}
 \left.\begin{array}{cc}{
 \bf {e^{1}}}\\  {\bf{e^{2}}}
\end{array}\right\}
= \frac{1}{\sqrt{2}}\,\left[\frac{\rho}{\Sigma} \,{\bf d \theta}\pm
i\,\frac{a\sin{\theta}\,\Sigma}{\rho}\left( {\bf{dt}}
-\frac{a^2+r^2}{a} {\bf d \phi} \right)\right],
\end{displaymath} and
\begin{displaymath}
 \left.\begin{array}{cc}{
 \bf {e^{3}}}\\  {\bf{e^{4}}}
\end{array}\right\}
=\frac{1}{\sqrt{2}}\,\left[\frac{\sqrt{Q}}{\rho}\left( {\bf{dt}} -
a\sin^2{\theta}{\bf d \phi} \right)\pm\,\frac{\rho} {\sqrt{Q}}\,{\bf
d r}\right],
\end{displaymath}
where $\Sigma={1-\frac{\Lambda}{3}a^2\,\cos^2{\theta}}$, it allows
to write down the metric as
\begin{eqnarray}
g=2{\bf{e^1}\bf{e^2}}-2{\bf{e^3}\bf{e^4}}=g_{ab}{\bf{e^a}\bf{e^b}},\,\,\,
\bf{e^a}&&={e^a}_\mu \,\bf{dx}^{\mu}.
\end{eqnarray}
Additionally to the null tetrad one may introduce an orthonormal
basis $\{\bf {E^a},a=1,\ldots,4\}=\{{\bf x},{\bf y},{\bf z},{\bf
t}\}$,
 where ${\bf t}$ is a time--like vector, ${\bf t\cdot t}=-1$,
such that ${\bf e^1}=({\bf x}+i{\bf y} )/\sqrt{2}$, ${\bf e^2}=({\bf
x}-i{\bf y} )/\sqrt{2}$, ${\bf e^3}=({\bf t}+{\bf z} )/\sqrt{2}$,
${\bf e^4}=({\bf t}-{\bf z} )/\sqrt{2}$. In coordinates
$\{\theta,r,\phi,t\}$, as this basis one defines:
\begin{eqnarray}&&\label{orthob}
{\bf E^1}=\frac{\rho}{\Sigma} \,{\bf d \theta}, \, {\bf
E^3}=\frac{a\sin{\theta}{\Sigma}}{\rho}\left( {\bf{dt}}
-\frac{a^2+r^2}{a} {\bf d \phi}\right),\,\nonumber\\&&
 {\bf E^2}=\frac{\rho}{\sqrt{Q}} \,{\bf d r},\,
 {\bf E^4}=\frac{\sqrt{Q}}{\rho}\left( {\bf{dt}} - a\sin^2{\theta}{\bf d
\phi} \right).
\end{eqnarray}
These bases are associated to the eigenvector bases; for instance;$
F_{\mu\nu}= 2F_{ab}{e^a}_{[\mu }\,{e^b}_{\nu]}=2F_{12}{e^1}_{[\mu
}\,{e^2}_{\nu]}+2F_{34}{e^3}_{[\mu }\,{e^4}_{\nu]}$ and
${\mathcal{E}^{\mu}}_{\nu}= \mathcal{E}^1_1\,{E^1}^{\mu
}\,{E^1}_{\nu }+ \mathcal{E}^2_2\,{E^2}^{\mu }\,{E^2}_{\nu }+
\mathcal{E}^3_3\,{E^3}^{\mu }\,{E^3}_{\nu }-
\mathcal{E}^4_4\,{E^4}^{\mu
}\,{E^4}_{\nu }$.\\
From the metric tensor one evaluates the coordinate components of
the Ricci tensor, the scalar curvature, and the Riemann-Weyl
curvature tensor. In particular, these curvature quantities,
refereed to the above null tetrad, acquire their simplest
description--their eigenvector structure. To avoid confusion in the
use of indices, we denote the coordinate indices with Greek--Latin
symbols $\{x^\mu \}=\{\theta,r,\phi,t\}$, the tetrad components
${E^a}_b, E_{ab}, a,b=1,\ldots,4$ with a prefix $N$ for null tetrad
components $NE_{ab}$ and $O$ for orthonormal tetrad components
$OE_{ab}$, avoiding in this manner the use of parentheses or
tildes.\\
The Einstein tensor is determined by a diagonal tensor matrix with
two pair of eigenvalue components, the corresponding traceless Ricci
tensor is described by a diagonal matrix too with two pair of
opposite in sign eigenvalues,
$\{\lambda_1,\lambda_1,-\lambda_1,-\lambda_1\}$, and hence,
according with the Segr\'e--Pleba\'nski classification of the
matter--field tensors, it can only describes (linear Maxwell and
nonlinear) electrodynamics  with non--zero invariants.\\

\section{Alignment conditions}

The alignment of the tetrad and eigenvectors of the field is,
perhaps, the very break point in the derivation of the
electromagnetic fields, and consequently of the resolution of the
entire problem.\\ The field tensor $F_{\mu\nu}$ is endowed with four
components: $ F_{\theta\phi}$,
 $F_{\theta\,t}$, $ F_{r\phi}$, $F_{rt} $; one looks for the eigenvectors $V_a^\mu$,  $a =1,\ldots,4$, of the
 tensor $F_{\mu\nu}$ by solving the corresponding  eigenvalue
 problem; the alignment of the eigenvectors $V_a^{\nu}$ along the
 tetrad basis, or equivalently, aligning the tetrad along the
 eigenvectors $V_a^{\nu}$ gives rise to the alignment conditions:
 \begin{eqnarray}\label{FieldFot1}&&
 {F_{r\phi}} =-a \, \sin^{2} { \theta }
 { F_{rt}} ,\,\,{ F_{\theta\phi}}
 =-{\frac {
   {a}^{2}+{r}^{2}  }{a}}{ F_{\theta t}},
\end{eqnarray}
thus  two of the field components are independent, say  $F_{r\, t}$
and $F_{\theta t}$, while the remaining two ${F_{r\phi}} $ and
$F_{\theta\phi}$ are determined through the alignment conditions.
Since the field tensor ${F_{\mu\nu}}$ is a curl, it can be
determined from its representation in term of the vector potential
$A_\mu$, $F_{\mu\nu}=A_{\nu,\mu}-A_{\mu,\nu} $. The alignment
conditions can be integrated for the electromagnetic vector
components $A_t$ and $A_{\phi}$: replacing
$F_{\theta\phi}=A_{\phi,\theta}$, $F_{r\phi}=A_{\phi,r}$, $F_{r
t}=A_{t,r}$, $F_{\theta t}=A_{t,\theta}$ one arrives at
 \begin{eqnarray}&&\label{SysAphi}
[{\frac {\partial A_\phi }{\partial r}}
  +a \,\sin^{2} {\theta} {\frac {
\partial {\it A_t}}{\partial r}} =0,\,{\frac {\partial A_\phi}{\partial \theta}}
 +{\frac {    {a}^{2}+{r}^{2}
  }{a}} {\frac {\partial  {\it A_t} }{\partial
\theta}}=0],
\end{eqnarray}
the integrability of $A_\phi$ leads to
\begin{eqnarray}&&
\frac {\partial }{\partial r}\left( \frac{ {a}^{2}+{r}^{2} }
 {a}\,\frac {\partial
 {\it A_t}}{\partial \theta} \right)-\frac {\partial }{\partial \theta}\left(
a \, \sin^2 {\theta} \,{\frac {
\partial {\it A_t}}{\partial r}}\right) =0,
\end{eqnarray}
i,e., an equation for $A_t$
\begin{eqnarray}&&\label{potAt0}
 (r^2+ {a}^{2}
\cos^{2} {\theta })
 \frac {\partial ^{2} {\it At}}{\partial \theta \partial r}
  +2\,r\, \frac {\partial {\it
A_t} }{\partial \theta} -2\,{a}^{2}\sin {\theta} \cos {\theta}
 \frac {\partial {\it At}}{\partial r}
=0,\nonumber\\&&
\end{eqnarray}
The general solution is sought in the form $
A_t=\frac{Z(\theta,r)}{{\rho }^{2}},$ substituting this expression
into the equation (\ref{potAt0}), one gets $\frac{\partial
}{\partial r}\,\frac{\partial }{\partial
\theta}Z(\theta,r)=0\rightarrow Z(\theta,r)=Y(\theta)+X(r)$, where
$Y(\theta)$ and $X(r)$, at this level, are arbitrary integration
functions, consequently
\begin{eqnarray}\label{POTAt}&&
{ A_t} ={\frac {Y ({\theta}) +X
 ( r ) }{{r}^{2}+{a}^{2}  \cos^{2} {\theta}
  }}
 \end{eqnarray}
This solution for $A_t$ guarantees the  integrability  of
$A_{\phi}$, (\ref{SysAphi}), which occurs to be
\begin{eqnarray}\label{potAphi0}&&
A_\phi =-{\frac {a \sin^{2} {\theta } X \left( r \right)
}{{r}^{2}+{a}^{2}
 \cos^{2} { \theta }  }}-{\frac { \left(
{a}^{2}+{r}^{2}
 \right) Y ({\theta})}{a \left({r}^{2}+ {a}^{2} \cos^{2}
 {\theta} \right) }}.
\end{eqnarray}

\section{Field equations $(\sqrt{-g_m}P^{\mu\nu})_{,\nu}=0$}

The dual tensor  ${}^{\star} P_{\mu\nu}$, being a curl, is defined
in terms of the vector potential components ${}^{\star} P_\mu$ ,
similarly as the $F_{\mu\nu}$, namely $ {}^{\star}
P_{\mu\nu}={}^{\star} P_{\nu,\mu}-{}^{\star} P_{\mu,\nu}$. In
correspondence with the structure of $F_{\mu\nu}$, as independent
dual components of ${}^\star P_{\mu\,\nu}$ one has the following: $
{}^\star P_{\,\theta\,t}={}^{\star}P_{t\,,\theta}
,\,{{}^\star}P_{r\,t}={}^{\star}P_{t\,,r}$.

  The Maxwell field equations $(\sqrt{-g_m}P^{\mu\nu})_{,\nu}=0$ , for
$(\sqrt{-g_m}P^{\phi\theta})_{,\theta}+(\sqrt{-g_m}P^{\phi\,r})_{,r}=0$
, where $g_m$ is the coordinate metric determinant, yield
  \begin{eqnarray}\label{EQLFG} &&{\frac {-a\,\sin {\theta}
   F_{rt}  {\it L_G}  + F_{\theta t}
  {\it L_F}  }{a\sin
{ \theta } }}={\frac {\partial }{\partial r}}\,{}^{\star}P_t\left(
\theta,r \right) ,
 \nonumber\\&&
a\sin{\theta }  F_{rt}  {\it L_F} +F_{\theta t} {\it L_G} =-{\frac {
\partial }{\partial \theta}}\,{}^{\star}P_t \left( \theta,r \right)
,
\end{eqnarray}
 which can be solved for the
 derivatives  ${\it L_F}$ and ${\it L_G}$, namely
    \begin{eqnarray}\label{derLFG}&&
{\it L_F} ={\frac {a\sin { \theta
 } \left( {F_{\theta\,t}} {\frac {
\partial {}^{\star}P_t }{\partial r}} -{F_{r\,t}}
 {\frac {\partial {}^{\star}P_t}{\partial \theta}} \right) }{ \left( {F_{r\,t}}
 \right) ^{2}{a}^{2}  \sin^{2} {\theta}   +
 \left( {F_{\theta\,t}}  \right) ^{2}}},
\nonumber\\&&
 {\it
L_G} =-{\frac {{ {a }^{2} F_{r\,t}} \,\sin^{2} {\theta}\,\,
 { \frac {\partial {}^\star P_t}{\partial r}}
+{F_{\theta\,t}} {\frac {\partial {}^\star P_t}{
\partial \theta}} }{ \left( {F_{r\,t}}
  \right) ^{2}{a}^{2}  \sin^{2} {
\theta }  + \left( {F_{\theta\,t}}   \right) ^{2}}},
\end{eqnarray}
which play an important role in the integration of the entire
problem.
 \\
While
$(\sqrt{-g_m}P^{t\theta})_{,\theta}+(\sqrt{-g_m}P^{t\,r})_{,r}=0$
gives rise to  alignment conditions for the tensor field  ${}^\star
P_{\mu\nu}$,
       \begin{eqnarray}&&\label{FieldPotP1}
 {}^\star P_{ r \phi}=-a
\sin^2{\theta} {}^\star P_{r\,t},\nonumber\\&&
 {}^\star P _{
\theta\,\phi}=-\frac{a^2+r^2}{a}\, {}^\star P_{\theta\,t} ,
\end{eqnarray}
which in terms of the vector potential components ${}^\star P_{\mu}$
read
\begin{eqnarray}&&\label{potPt0}
\frac{\partial}{\partial r}  {}^\star P_{\phi}=-a
\sin^2{\theta}\frac{\partial}{\partial r} {}^\star P_t,\nonumber\\&&
\frac{\partial}{\partial {\theta}} {}^\star P
_{\phi}=-\frac{a^2+r^2}{a}\frac{\partial}{\partial \theta} {}^\star
P_t,
\end{eqnarray}
which are similar in all respect to the equations for
 $A_t$ and $A_\phi$ , (\ref{SysAphi}), therefore the solutions for
${}^{\star}P_{t}$ and  ${}^{\star}P_{\phi}$  are:
\begin{eqnarray}\label{PotPt1}&&
{}^\star P_{t}={\frac {A \left( r \right) +B \left( \theta \right)
}{{a}^{2} \left( \cos {\theta}  \right) ^{2}+{r}^{2}}} ,
\end{eqnarray}
\begin{eqnarray}\label{potPphi0}&&
-{}^{\star}P_{\phi}={\frac {a \left( \sin
 {\theta}  \right) ^{2}A \left( r \right) }{{a}^{2}
 {\cos^2{\theta}}+{r}^{2}}}+{\frac {
 \left( {a}^{2}+{r}^{2} \right) B (\theta) }{a \left( {a}
^{2} {\cos^2{\theta}}+{r}^{2} \right) } }.
\end{eqnarray}

\section{The KEY equation}
The expressions (\ref{derLFG}) of the derivatives  ${\it L_F}$ and
${\it L_G}$, can be used to calculate the derivatives ${\it L_{r}}$
and ${\it L_{\theta}}$ of the Lagrangian $L$:
\begin{eqnarray}\label{derLFG2}&&
\frac{\partial {L} }{\partial r}=L_F \frac{\partial {F} }{\partial
r}+L_G \frac{\partial {G} }{\partial r},\nonumber\\&& \frac{\partial
{L} }{\partial {\theta}}=L_F \frac{\partial {F} }{\partial
{\theta}}+L_G \frac{\partial {G} }{\partial {\theta}}
\end{eqnarray}
Next, using the expressions of the invariant $F$ and $G$ under the
alignment conditions (ac) (\ref{FieldFot1}) and  (\ref{FieldPotP1})
one obtains
\begin{eqnarray}\label{InvFGc}&&
{\it InvF_{ac}}=\frac{1}{2}\,\frac {{a}^{2}
\sin^{2}{\theta}\,{F_{\theta t}}^{2}- {F_{rt}}^2 }{
a^2\,\sin^{2}{\theta} }, \nonumber\\&& {\it InvG_{ac}}=-{\frac
{{F_{\theta\,t}} {F_{r\,t}}
 }{a\sin {\theta} }}
\end{eqnarray}
together with the substitution of $L_F$ and $L_G$  from
(\ref{derLFG})  lead to the simple expressions
\begin{eqnarray}\label{derLr}&&
{\frac {\partial }{\partial r}}L = {\frac { \left( {\frac {\partial
}{\partial r}}{}^{\star}P_t \right) {\frac {\partial }{\partial
r}}{\it F_{\theta t}} + \left( {\frac {\partial }{\partial
\theta}}{}^{\star}P_t
  \right) {\frac {\partial }{\partial r}}{\it
F_{rt}} }{a\sin {\theta} }}
\end{eqnarray}
\begin{eqnarray}\label{derLteta}&&
{\frac {\partial }{\partial \theta}}L=-{\frac {{F_{\theta\,t}}\,\cos
{\theta} \,  \left( { \frac {\partial }{\partial r}}{}^{\star}P_t
\right) }{a \,\sin^{2}{ \theta }
  }}
 \nonumber\\&&+{\frac { \left( {\frac {\partial }{\partial \theta}}{
\it F_{rt} } \right) {\frac {\partial }{\partial
\theta}}{}^{\star}P_t }{a\sin {\theta} }}+ {\frac { \left( {\frac
{\partial }{\partial \theta}}{\it F_{\theta t}}
  \right) {\frac {\partial }{\partial r}}{}^{\star}P_t
 }{a\sin {\theta} }} .
\end{eqnarray}

 The integrability  of the Lagrangian 1--form equation
 \begin{eqnarray}\label{Lagrangian}&&
 dL(\theta,r) =\frac{\partial L}{\partial \theta}\,d\,\theta
 +\frac{\partial L}{\partial r}\,d\,r
     \end{eqnarray}
is guaranteed by the closure condition $d^2\,L=0$, i.e., from
vanishing of the mixed derivative equation $\frac{\partial}{\partial
\theta}\frac{\partial L}{\partial r}=\frac{\partial}{\partial
r}\frac{\partial L}{\partial \theta}$, replacing $F_{rt}=A_{t,r}$,
$F_{\theta t}=A_{t,\theta}$, one arrives at the key equation
\begin{eqnarray}&&
KEY:=\frac{\partial^2 A_t}{\partial {r}\,\partial r}
\cdot\,\frac{\partial }{\partial \theta}\left(
\frac{1}{\sin{\theta}}\frac{\partial {}^{\star}P_t}{\partial
\theta}\right)-\frac{\partial^2{{}^{\star}P_t}}{\partial
{r}\,\partial r} \,\cdot\frac{\partial }{\partial \theta}\left(
\frac{1}{\sin{\theta}}\frac{\partial A_t}{\partial \theta}\right)=0.
\end{eqnarray}

In the search for solutions of the Carter--Pleba\'nski metric, which
will be published elsewhere soon, using the coordinates
$x=a\cos{\theta}\,,\,y=r$ this KEY equation is simply
\begin{eqnarray}&&
 KEY:=\frac{\partial^2 A_t}{\partial {y}^2}
\cdot\,\frac{\partial^2 {}^{\star}P_t}{\partial
x^2}-\frac{\partial^2 A_t }{\partial
x^2}\,\cdot\frac{\partial^2{}^{\star}P_t}{\partial {y}^2}
 =0 .
\end{eqnarray}
The approach to derive solutions of KEY is through the integrals of
$A_t$ and ${}^{\star}P_t$; substituting them into the KEY one
arrives at single nonlinear second order equation for the potential
functions $\{Y(\theta),X(r),A(r),B(\theta)\}$, one may proceed
further by using a separation of variables procedure.  Any solution
of this KEY equation ensures the integrability of the Lagrangian
function via the equation (\ref{Lagrangian}).

\subsection{ Vector potentials of the Kerr--Newman solution}\label{VectPotKN}

The simplest solution of the KEY equation corresponds to the
Kerr--Newman solution since its vector potential components $A_t$
and ${}^{\star}P_t$ fulfil the Laplace equation
\begin{eqnarray}\label{Laplace}&&
 \frac{\partial^2 A_t}{\partial {y}^2}
+\frac{\partial^2 A_t }{\partial x^2}\,=0 \rightarrow
\frac{\partial^2 {}^{\star}P_t}{\partial
x^2}+\frac{\partial^2{}^{\star}P_t}{\partial {y}^2} ,\,\,KEY\equiv
0.
\end{eqnarray}
 Therefore, their solutions are of the form
\begin{eqnarray}&&
 H_t=H(x+i\,y)+ \bar{H}(x-i\,y).
\end{eqnarray}
  For the solution $2\,H(x+i\,y)=\frac {a_0+i\,b_0}{x+i\,y}$
\begin{eqnarray}&&
2\,H_t(x,y)= \frac {a_0+i\,b_0}{x+i\,y} +\frac
{a_0-i\,b_0}{x-i\,y}=2\frac{a_0\,x+b_0\,y}{x^2+y^2}\rightarrow
{H_t(\theta,r)=\frac{a_0\,a\,\cos{\theta}+b_0\,r}{a^2\cos^2{\theta}+r^2}}
,
\end{eqnarray}
 hence
 \begin{eqnarray}&&
{A_t(\theta,r)=\frac{a_0\,a\,\cos{\theta}+b_0\,r}{a^2\cos^2{\theta}+r^2}},
\end{eqnarray}
which is just the electromagnetic vector potential $A_t$ for the
Kerr--Newman metric by identifying $b_0=e,\,\,a_0=g_0$ which are the
electric $e$ and magnetic $g_0$ charges; the field component
$A_\phi$ is obtained from the the alignments conditions
(\ref{FieldFot1}); it results in
\begin{eqnarray}&&\label{PotenAkn}
-\,A_\phi= \frac{
 {\it a_0}\, \left( {a}^{2}+{r}^{2} \right)\cos{\theta} + a{\it
b_0}\,r \, \sin ^{2}{\theta}}{a^2\cos^2{\theta}+r^2}.
\end{eqnarray}
 For the components ${}^{\star}P_t$ and
${}^{\star}P_\phi$ one obtains similar expressions;
$A_t(a_0\rightarrow f_0,b_0\rightarrow h_0)\rightarrow {}^{\star}P_t
$, and $A_\phi(a_0\rightarrow f_0,b_0\rightarrow h_0)\rightarrow
{}^{\star}P_\phi $, where $f_0$ and $h_0$ are constants related with
the charges. The field tensors $F_{\mu\nu}$ and
${}^{\star}P_{\mu\nu}$ are derived by differentiation of these
vector potentials $A_\mu$ and ${}^{\star}P_{\mu}$.

Of course one could directly derive the electromagnetic fields of
the Kerr-Newman solution integrating the (\ref{Faraday}) and
(\ref{Maxwell}) equations, which under the alignment conditions
 become (\ref{EQLFG}); for the Maxwell electrodynamics $L=F,\,
L_F=1,\,L_G=0$, the equations (\ref{EQLFG}) reduces to
 \begin{eqnarray}&&\label{PotenAkn2}
    \frac{1}{a\sin{\theta}}F_{\theta t}
    =\frac{\partial}{\partial r} {}^{\star} P_t
 \rightarrow{\frac{1}{a\sin{\theta}}\frac{\partial}{\partial \theta} A_t
 = \frac{\partial}{\partial r} {}^{\star} P_t
 } \nonumber\\&&
  a\, \sin{\theta}F_{rt} =-\frac{\partial}{\partial \theta} {}^{\star} P_t
 \rightarrow{\frac{\partial}{\partial r} A_t
 =-\frac{1}{a\,\sin{\theta}} \frac{\partial}{\partial \theta} {}^{\star} P_t
 },
\end{eqnarray}
 which in terms of the variables $x=a\sin{\theta}$ and $y=r$ rewrite
 as
     \begin{eqnarray}&&\label{PotenAkn2}
\frac{\partial}{\partial x} A_t
 = \frac{\partial}{\partial y}{}^{\star} P_t,
 \,\frac{\partial}{\partial y} A_t
 = -\frac{\partial}{\partial x}
{}^{\star} P_t
 ,
\end{eqnarray}
with integrability conditions:
   \begin{eqnarray}&&\label{PotenAkn2}
A_{t,y,x}=A_{t,x,y}\rightarrow {}^{\star} P_{t,x,x}+{}^{\star}
P_{t,y,y}=0 , \nonumber\\&&
  {}^{\star}P_{t,y,x}={}^{\star}P_{t,x,y}\rightarrow  A_{t,x,x}+
A_{t,y,y}=0 ,
\end{eqnarray}
i.e., the starting point (\ref{Laplace}) of the integration process
we followed above, \ref{VectPotKN}. It is quite possible that the
Kerr--Newman solution is singled out as the unique stationary
axisymmetric solution whose electromagnetic potentials are derivable
in Einstein--NLE from a Lagrangian function depending
explicitly on the $F$ and $G$, $L(F,G)= A_0 F+B_0 G.$, see \ref{LagrangianKN}.\\

As we show in the forthcoming sections, the ``straightforward
coordinate dependent method'' in nonlinear electrodynamics
(SCDM--NLE) permits the determination, from electromagnetic field
tensors $F_{\mu\nu}(x^\alpha)$ and ${}^{\star}P_{\mu\nu}(x^\alpha)$
fulfilling the conservation ``Faraday--Maxwell'' equations, the
corresponding coordinate dependent Lagrangian function
$L(x^\alpha)$.  In this ``method'' one does not need to know nor
establish the explicit dependence of the Lagrangian on the the
invariants $F$ and $G$; these quantities as result of their
definition through the electromagnetic field tensor depend on the
coordinate too.

\section{The cubic vector potentials}\label{cubivcsol}

The cubic potential functions $\{A(r),B(\theta),X(r),Y(\theta)\}$,
 of the form $  F(r)=f_0+f_1r+f_2 x^2+ f_3 r^3$, and $G(\theta)=
 g_0+g_1 \cos{\theta}+ g_2 \cos^2{\theta}  +g_3\cos^3{\theta} $,
 substituted in KEY gives rise to a polynomial in $r$ and
 $\cos{\theta}$, equating to zero the coefficients of the independent
 powers $r^s\,\cos^m{\theta}$ one gets an algebraic system of
 equations, whose solutions determines the cubic, in this example,
 potential functions
\begin{eqnarray}&&
 A( r) =-{r}^{3}{\it y_1}\,\beta+ {\it y_1}\,r,
\nonumber\\&&B(
 {\theta}) ={a}^{3} {\it x_1}\,\beta\, \cos^{3}{
\theta }+a {\it x_1}\,\cos {\theta} ,\nonumber\\&&X({\theta})
={a}{\it f_1} \,\beta\,\cos^{3} { \theta} +a {\it f_1}\,\cos
{\theta} ,\nonumber\\&&Y(r)= -{r}^{3} {\it g_1}\,\beta+ {\it
g_1}\,r,
\end{eqnarray}
which, in turn, determine via (\ref{POTAt}) and (\ref{potAphi0}) the
vector potential components of the vector field
$A_{\mu}=\{0,0,A{_\phi}, A{_t}\}$:
\begin{eqnarray}&&\label{PotenA}
\rho^2\, A_{t}= \left({\it f_1}\,{a}^{3}  \cos ^{3} \theta
 -{\it g_1}\,{r}^{3} \right) \beta
+{{ \it f_1}\,a\cos{\theta}  +{\it g_1}\,r}, \nonumber\\&&
-\rho^2\,A_\phi=a\,\beta \left[  {a}{\it f_1}\, \left(
{a}^{2}+{r}^{2} \right) \cos^{3} {\theta} -{\it g_1}\,{r}^{3}
\sin^{2} { \theta }
 \right]
 \nonumber\\&&+
 {\it f_1}\, \left( {a}^{2}+{r}^{2} \right)\cos{\theta} + a{\it
g_1}\,r \, \sin ^{2}{\theta}.
\end{eqnarray}
The dual vector potential ${}^{\star} P_\mu$ is given, via
(\ref{PotPt1}) and (\ref{potPphi0}), by ${}^{\star}P_t$ and
${}^{\star}P_\phi$:
\begin{eqnarray}&&\label{PotenP}
\rho^2\,{}^{\star}P_t={\beta}({a}^{3}\,{\it x_1}\, \cos^{3} {\theta}
-{\it y_1}\,{r}^{3})+ a {\it x_1}\,\cos {\theta}+{\it y_1}\,r,
\nonumber\\&& -\rho^2\,{}^{\star}P_\phi= {a} {\beta}\left[
a\,x_1\,\left( {a}^{2}+{r}^{2} \right)\, \cos^{3}{\theta}
 \,  -  {\it y_1}{r}^{3}\, \sin^{2}
{\theta} \right] \nonumber\\&&+ {\it x_1}\,(r^2+a^2)\cos{ \theta
}+a{\it y_1}\,r\,\sin^{2} {\theta}.
\end{eqnarray}
Accomplishing the corresponding partial differentiation of the
components $A_t$, and $A_\phi$ one gets the field tensor components
$F_{\mu\nu}$ and ${}^{\star}P_{\mu\nu}$:
\begin{eqnarray}&&
F_{rt}=-{\frac {r\beta\, \left( 2\, {\it f_1}\,{a}^{3}\,\cos^{3}
 {\theta} +3\,{a}^{2} {\it g_1}\,r \cos^{2} { \theta} + {\it g_1}\,{r}^{3} \right) }{
 \left( {a}^{2} \cos^{2} {\theta} +{r}^{2
} \right) ^{2}}} \nonumber\\&&+{\frac { {\it g_1}\, \left( {a}^{2}
 \cos^{2} {\theta} -{r}^{2} \right) -2\, a {\it f_1}\,r\,\cos {\theta
}}{ \left( {a}^{2}  \cos^{2}{\theta}
 +{r}^{2} \right) ^{2}}} ,
\end{eqnarray}

\begin{eqnarray}&&
F_{\theta t}=-{\frac {\beta\,{a}^{2}\cos {\theta} \sin
 {\theta} \, \left[ {\it f_1}\,a\cos { \theta
}  \left( {a}^{2}  \cos^{ 2} {\theta} +3\,{r}^{2} \right)+2\,{\it
g_1}\,{r}^{3} \right] }{ \left( {a}^{2}
 \cos^{2} {\theta} +{r}^{2} \right) ^{2}}
}\nonumber\\&&+{\frac {a\sin {\theta}  \left[ {\it f_1}\, \left(
{a}^{2 } \cos^{2} {\theta} -{r}^{2} \right) +2\,a {\it g_1}\,r\,
\cos {\theta}  \right]}{ \left( {a}^{2}
  \cos^{2} {\theta}  +{r}^{2} \right) ^{2}}},
\end{eqnarray}
while the remaining field components are evaluated from the
alignment conditions
\begin{eqnarray}&&
F_{\theta\,\phi\,}=-{\frac { {a}^{2}+{r}^{2} }{a}}\, F_{\theta\,t},
\nonumber\\&& F_{\,r\,\phi}=-a \, \sin^{2} {\theta } \, F_{r\,t} .
\end{eqnarray}
On the other hand, accomplishing the corresponding differentiation
of the vector  potential components ${}^{\star}P_t$ and
${}^{\star}P_\phi$  one gets the tensor field components
${}^{\star}P_{\mu\nu}$:
\begin{eqnarray}&&
{}^\star P_{rt}=-{\frac {r\beta\, \left( 2\,{\it
x_1}\,{a}^{3}\,\cos^{3}
 {\theta} +3\,{a}^{2} {\it y_1}\,r \cos^{2} { \theta} + {\it y_1}\,{r}^{3} \right) }{
 \left( {a}^{2} \cos^{2} {\theta} +{r}^{2
} \right) ^{2}}} \nonumber\\&&+{\frac { {\it y_1}\, \left( {a}^{2}
 \cos^{2} {\theta} -{r}^{2} \right) -2\, a {\it x_1}\,r\,\cos {\theta
}}{ \left( {a}^{2}  \cos^{2}{\theta}
 +{r}^{2} \right) ^{2}}}  ,
\end{eqnarray}

\begin{eqnarray}&&
{}^\star P_{\theta t}=-{\frac {\beta\,{a}^{2}\cos {\theta} \sin
 {\theta} \, \left[ {\it x_1}\,a\cos { \theta
}  \left( {a}^{2}  \cos^{ 2} {\theta} +3\,{r}^{2} \right)+2\,{\it
y_1}\,{r}^{3} \right] }{ \left( {a}^{2}
 \cos^{2} {\theta} +{r}^{2} \right) ^{2}}
}\nonumber\\&&+{\frac {a\sin {\theta}  \left[ {\it x_1}\, \left(
{a}^{2 } \cos^{2} {\theta} -{r}^{2} \right) +2\,a {\it y_1}\,r\,
\cos {\theta}  \right]}{ \left( {a}^{2}
  \cos^{2} {\theta}  +{r}^{2} \right) ^{2}}} ,
\end{eqnarray}
the remaining two components are evaluated from the corresponding
alignment conditions
\begin{eqnarray}&&
{}^{\star}P_{\theta\,\phi\,}=-{\frac { {a}^{2}+{r}^{2} }{a}}{}^\star
P_{\theta\,t}, \nonumber\\&& {}^{\star}P_{\,r\,\phi}=-a \, \sin^{2}
{\theta } \,{}^\star P_{r\,t}.
\end{eqnarray}

Notice the symmetry between these tensor field components,
$$\{A_t,\,A_\phi,\,F_{\theta\,\phi\,},F_{\theta\,t\,},F_{r\phi}, F_{rt}\}\leftrightarrows
\{{}^{\star}P_{t},\,{}^{\star}P_{\phi},\,
{}^{\star}P_{\theta\,\phi\,}, {}^{\star}P_{\theta\,t\,},
{}^{\star}P_{r\,\phi\,}, {}^{\star}P_{r\,t\,}\}, $$ under the
replacement $f_1\leftrightarrows x_1$ and $g_1\leftrightarrows y_1$.

\section{Lagrangian function for the cubic solution} \label{LagrangianKN}

To get the Lagrangian function $L$, one integrates the 1--form $dL=
\frac{\partial L}{ \partial r}\,d\,r  + \frac{\partial L}{
\partial \theta}\,d\,{\theta} $, (\ref{Lagrangian}), where the
derivatives $ \frac{\partial L}{ \partial r}$ and $\frac{\partial
L}{
\partial \theta} $, (\ref{derLr}) and (\ref{derLteta}), are
evaluated for the cubic vector potentials and tensor field
components  derived in the previous section {\ref{cubivcsol}}. After
a time consuming integration, one can give the Lagrangian function
in the form
\begin{eqnarray}&&
L=\frac{1}{2\,\rho^8\,}\left(L_{\beta^2}\beta^2+L_{\beta}\beta+L_{\beta^0}
\right)+L_0
\end{eqnarray}
where $L_0$ is an integration constant and the functions
$L_{\beta^i}, \,i=0,1,2$ denote the Lagrangian numerators given
correspondingly as
\begin{eqnarray}&&
L_{\beta^2}=4\,  \cos^{7} {\theta}  {a}^{7} {\it f_1}\,r{\it
x_1}+6\, \cos ^{6 } {\theta}{a}^{6}{\it g_1}\,{r}^{2}{\it x_1}
 +12\,\cos^{5} {\theta
}  {a}^{5}{\it f_1}\,{r}^{3}{\it x_1}\nonumber\\&&+5\, \cos^{4}
 {\theta}{a}^{4}{\it f_1}\,{r}^{4}{\it y_1}
 +
23\,  \cos^{4} {\theta} {a}^{4}{\it g_1}\,{ r}^{4}{\it x_1} +12\,
\cos^{3} {\theta} {a} ^{3}{\it g_1}\,{r}^{5}{\it y_1}
 \nonumber\\&&
-6\,\cos^{2}{\theta}
 {a}^{2}{\it f_1}\,{r}^{6}{\it y_1}
 +12\,\cos^{2} {
\theta } {a}^{2}{\it g_1}\,{r}^{6}{\it x_1}+4\,\cos
 {\theta} a{\it g_1}\,{r}^{7}{\it y_1}
\nonumber\\&&
 -3\,{\it f_1}\,{r}^{8}
{\it y_1}+3\,{\it g_1}\,{r}^{8}{\it x_1},
\end{eqnarray}
\begin{eqnarray}&&{L_{\beta}}=2\,{\it f_1}\,{r}^{6}{\it y_1}+ 6\,{a}^{2}\left( 2{\it
f_1}\, {\it y_1}
 +{\it g_1}\,{\it x_1} \right)
{r}^{4}\,{\cos^2{\theta}}\nonumber\\&&+16\,{a}^{3} \left( {\it
f_1}\,{\it x_1} -{\it g_1}\,{\it y_1} \right) {r}^{3} \cos^{3} {
\theta} \nonumber\\&&-6\,{a}^{4} \left( {\it f_1}\,{\it y_1} +2{\it
g_1}\,{\it x_1} \right) {r}^{2}  \cos^{4} { \theta
 }-2\,\cos^{6} {\theta}
{a}^{6}{\it g_1}\,{\it x_1} ,
\end{eqnarray}
\begin{eqnarray}\label{LagrangianFree}&&
L_{\beta^0}= \left( {a}^{4}  \cos^{4} {\theta}
  -6a^2r^2 \cos^2{\theta}+{r}^{4} \right)  \left( {\it f_1}\,{\it y_1}+{\it g_1}\,{
\it x_1}\right) \nonumber\\&& -4 \,a r\,\cos {\theta}  \left(
a^2\cos^2{\theta}-r^2
 \right)  \left( {\it
f_1}\,{\it x_1}-{\it g_1}\,{\it y_1} \right)  ,
\end{eqnarray}
this last expression $L_{\beta^0}$ corresponds to the numerator of
the Maxwell Lagrangian  $L_{\beta^0}=L_{Maxwell}$.\\

\subsection{The Lagrangian $L_{\beta^0}$ in terms of
the invariants at zero order of beta}

   The evaluation of the invariants (\ref{InvFGc}) for
   the cubic solution at zero degree of $\beta$ gives:\\
\begin{eqnarray}&&
InvG_{\beta^0}=\frac{2}{\rho^8} \, a\,r\,
 \cos \left( \theta \right) \left( a^2\,\cos^2 {\theta}
 -r^2 \right)\left({\it f_1
} ^2-{\it g_1 }^2 \right) \nonumber\\&&- \frac{1}{\rho^8} \left(
{a}^{4} \cos^{4} {\theta} -6\,{a}^{2}\,{r} ^{2}\,
\cos^{2}{\theta}+{r}^{4} \right) {\it f_1 }{\it g_1}.
\end{eqnarray}
while the invariant $InvF_{\beta^0}$ occurs to be
\begin{eqnarray}&&
InvF_{\beta^0}=\frac{1}{2\rho^8}\left( {a}^{4} \cos^{4} {\theta}
-6\,{a}^{2}\,{r} ^{2}\,  \cos^{2}{\theta}+{r}^{4} \right) ({{\it
f_1}}^{2}-{{\it g_1}}^{2})\nonumber\\&&+ \frac{4}{\rho^2} \, a\,r\,
\cos {\theta} \left( a^2\,\cos^2 {\theta}
 -r^2 \right)\,
{\it g_1}\,{\it f_1}.
\end{eqnarray}
Thus one can define the auxiliary functions $Z$ and $W$
\begin{eqnarray}
{\it Z}=  {a}^{4}\,\cos^{4} {\theta}-6\,{a}^{2}{r}^{2}\,
 \cos^{2}{\theta}+{r}^{4},
\end{eqnarray}
\begin{eqnarray}
{\it W}=a\,\cos {\theta }  \left( {a}^{2} \, \cos^{2}
 {\theta }-{r}^{2} \right) ,
\end{eqnarray}
and build the algebraic system for:
\begin{eqnarray}
{\it I}=2\rho^8 InvF_{\beta^0},\,\, {\it J}=\rho^8 InvG_{\beta^0},
\end{eqnarray}
namely
\begin{eqnarray}
{\it I}={\it Z}\,{\it F_1}+8\,{\it W}\,{ \it G_1},\,\,{\it
J}=2\,{\it W}\,{\it F_1}-{\it Z}\,{\it G_1},
\end{eqnarray}
where
\begin{eqnarray}
 {\it F_1}={{\it f_1}}^{2}-{{\it g_1}}^{2},{\it G_1}={\it g_1}\,{\it
 f_1}.
 \end{eqnarray}
Solving for $W$  and $Z$ one gets
 \begin{eqnarray} \label{FuncWZ}
 {\it W}=\,\frac{{\it F_1}\,{\it J
}+{\it G_1}\,{\it I}}{2({{\it F_1}}^{2}+4\,{{\it G_1}}^{2})} ,{\it
Z}={\frac {{\it F_1}\,{\it I}-4\,{\it G_1}\,{\it J}}{{{\it
F_1}}^{2}+4\,{{\it G_1}}^{2}}} .
\end{eqnarray}
The Lagrangian $L_{\beta^0}$, from (\ref{LagrangianFree}), in terms
of $W$ and $Z$ becomes
\begin{eqnarray}
  2\,\rho^8\,{\it L_M}={\it Z}\, \left( {\it f_1}\,{\it y_1}+{\it g_1}\,{\it x_1}
 \right) -4\,{\it W}\, \left( {\it f_1}\,{\it x_1}-{\it g_1}\,{\it y_1}
 \right),
 \end{eqnarray}
substituting $W$ and $Z$, from (\ref{FuncWZ}), leads to $L_M$ in
terms of the invariants
\begin{eqnarray}
 2\,\rho^8\,{\it L_M}={\frac {{\it F_1}\,{\it I}-4\,{\it G_1}\,{\it J}}{{{\it
F_1}}^{2}+4\,{{\it G_1}}^{2}}}\, \left( {\it f_1}\,{\it y_1}+{\it
g_1}\,{\it x_1}
 \right) -2\,\,\frac{{\it F_1}\,{\it J
}+{\it G_1}\,{\it I}}{{{\it F_1}}^{2}+4\,{{\it G_1}}^{2}}\, \left(
{\it f_1}\,{\it x_1}-{\it g_1}\,{\it y_1}
 \right),
 \end{eqnarray}
gathering the invariants ${\it I}$ and ${\it J}$ and substituting
$F_1$ and $G_1$, one arrives at
\begin{eqnarray}&&
 2\rho^2 \,L_M=I\frac{f_1\,y_1-g_1\,x_1}{f_1^2+g_1^2}  -J \frac{f_1\,x_1+g_1\,y_1}{f_1^2+g_1^2}
\nonumber\\&&=I\frac{F_0}{f_1^2+g_1^2}  -J
\frac{G_0}{f_1^2+g_1^2}=2\rho^8\,InvF_{\beta^0}\frac{F_0}{f_1^2+g_1^2}
-\rho^8\,InvG_{\beta^0}\frac{G_0}{f_1^2+g_1^2},
 \end{eqnarray}
or
  \begin{eqnarray}
 L_{Max}=  =2\,InvF_{\beta^0}\,\frac{F_0}{f_1^2+g_1^2}
-\,InvG_{\beta^0}\,\frac{G_0}{f_1^2+g_1^2}.
 \end{eqnarray}
The  Maxwell relation $L=F$ arises for
\begin{eqnarray}&&f_1=g_0=y_1\,\, \text{and}\,\,
g_1=e=-x_1,\,\,\text{then} \,\,\nonumber\\&&
F_0:=f_1\,y_1-g_1\,x_1\rightarrow e^2+g_0^2 \,\,\text{and} \,\,
G_0:= f_1\,x_1+g_1\,y_1  \rightarrow 0, \,\text{hence} \nonumber\\&&
 L_{\beta^0}=InvF_{\beta^0}=F .
\end{eqnarray}
Therefore, we have demonstrated that the general Lagrangian
$L(\beta)$ in the linear Maxwell limit becomes the Maxwell
Lagrangian $L=F$. \\

Since the function W and Z are polynomials of third
    and fourth degree in the coordinates $x=a\cos{\theta} $  and
    $r$ one can express this coordinates
    in terms of the Maxwell
    invariants $F_M$ and $G_M$, which are awful enough, as new coordinates and gets the
    Lagrangian L as function of $F_M$ and $G_M$.  The problem of solving for $r$
    and $x$ analytically in term of the whole invariants $F_{\beta}$ and $G_{\beta}$ faces
   the insuperable problem of searching the roots for
   polynomials (with radicals) of eight degree in $r$ and
    $x$; for the impossibility of a polynomial to be solvable by radicals see N.H. Abel
    and E. Galois works, thus the remaining alternative is the numerical analysis
    approach which is widely used in Mathematical Physics.

 \section{Integration of the Einstein equations }
 \marginpar{Check NULL}
The Einstein tensor ${NE^a}_b$ possesses two pairs of {\it
different} eigenvalues, ${NE^2}_2={NE^1}_1={NE}_{12}$ and
${NE^3}_3={NE^4}_4=-NE_{34}$, such that the tensor  matrix
$({NE^a}_b)=\text {diag}({NE^1}_1,{NE^1}_1,{NE^4}_4,{NE^4}_4)$ while
the orthonormal tetrad components fulfil
${OE^2}_2={OE^4}_4={OE}_{11}$ and ${OE^3}_3={OE^1}_1=OE_{44}$, with
matrix tensor
 $({OE^a}_b)=\text
{diag}({OE^1}_1,{OE^4}_4,{OE^1}_1,{OE^4}_4)$;
$(OE_{ab})=\text{diag}(OE_{11},-OE_{44},OE_{11},OE_{44})$.
There is
a relationship between these tetrad components

\begin{eqnarray}
 NE_{12} =  OE_{11},\, \,NE_{34} =  OE_{44} .
 \end{eqnarray}
We consider as independent Einstein equations those given by the
orthonormal components $OE_{11}$ and $OE_{44}$:
\begin{eqnarray}\label{OEDD11}&&
  OE_{11}=\frac{\Lambda\,a^4}{3\rho^4}
  \cos^2{\theta}\left(3\cos^2{\theta}-1\right)\nonumber\\&&
  +\frac{1}{2\rho^4 }  \left( \rho^2 {Q^{\prime\prime}}
  -2\,r\,Q^{\prime}+2 Q-2\,a^2\,\cos^2{\theta} -2\,a^2 \right)  \nonumber\\&&
  =\kappa\,OT_{11} +\Lambda \eta_{11}=\kappa\, \left(L-{\frac {  F_{\theta\,t}  {\frac {
\partial {}^{\star} P_t}{\partial r}}  }{a\sin {
\theta}}}\right)  +\Lambda \nonumber\\&&=\Lambda
+\frac{1}{2\,\rho^4}\left(\rho^2 {K^{\prime\prime}}
  -2\,r\,K^{\prime}+2 K\right),
\end{eqnarray}
and
\begin{eqnarray}\label{OEDD44}&&
    OE_{44} =  -\frac{\Lambda\,a^2}{3\rho^4}
 \left(3\,a^2\,\cos^4{\theta}  +6r^2\cos^2{\theta}
  -r^2  \right)-\frac{1}{\rho^4}\left(r\,Q^{\,\prime}-Q-r^2+a^2 \right)  \nonumber\\&&
  =\kappa\,OT_{44}+\Lambda \eta_{44} =\kappa\, \left(-L  +{\frac {
F_{r\,t} {\frac {\partial {}^{\star} P_t }{\partial \theta}}
 }{a\sin {\theta } }} \right)  -\Lambda \nonumber\\&& =-\Lambda-
 \frac{1}{\rho^4}\left(r\,  K^{\prime} -K
 \right) .
\end{eqnarray}
Adding these equations one arrives at the traceless Ricci tensor
eigenvalue $S=S^1_1$ equation:

\begin{eqnarray} &&
   2\,S^1_{1}-\kappa(OT_{11}+OT_{44}) =  -\frac{\Lambda\,a^2}{3\rho^4}
 \left(\,a^2\,\cos^2{\theta}  +6\,r^2\,\cos^2{\theta}
  -r^2  \right)\nonumber\\&&
  +\frac{1}{2\rho^4 }  \left( \rho^2 {Q^{\,\prime\prime}}
  -4\,r\,Q^{\,\prime}+4\, Q-2\,a^2\,\cos^2{\theta}-4\,a^2+2\,r^2  \right)
  \nonumber\\&& -\frac{\kappa}{a\sin{\theta}} \left(\left(\frac
{\partial {\it A_t}}{
\partial \theta}  \right) \frac {
\partial {}^{\star} P_t}{\partial r} - \left( \frac
{\partial {\it A_t} }{
\partial r} \right) \frac {
\partial {}^{\star} P_t}{\partial \theta}  \right) =:\frac{1}{6\rho^4}
\left(V(r)\,a^2\,\cos^2{\theta}+W(r)\right),
\end{eqnarray}
where the contribution of the electromagnetic field is represented
by
\begin{eqnarray}&&
OT_{44}+OT_{11}={\frac { \left( {\frac {\partial }{
\partial r}} {}^{\star} P_t  \right) {\frac {\partial }{
\partial \theta}}{\it A_t} - \left( {\frac {
\partial }{\partial \theta}}{}^{\star} P_t  \right) {
\frac {\partial }{\partial r}}{\it A_t} }{a\sin
 \left( \theta \right) }},
\end{eqnarray}
whose evaluation gives
\begin{eqnarray}&&
\kappa(OT_{11}+OT_{44})=\frac{\kappa\,F_0}{\rho^4} \left(
3\,{\beta}^{2}\,{r}^{2}\,{a}^{2} \cos^{2}
 {\theta }-{a}^{2}\,\beta\,
\cos^{2} {\theta } +\beta\,{r}^{2}+1
 \right) .
\end{eqnarray}

Therefore the auxiliary functions $V(r)$,and $W(r)$ which stand for
the separable equation terms are:\\the $V(r)$ equation is given by
\begin{eqnarray}&&
V(r)=3\,{\frac {d^{2}Q}{d{r}^{2}}}
 \,+6\,\beta\,\kappa\,
F_0 \left( 1-3\,\beta\,{r}^{2} \right) -2\,{a}^{2}\Lambda-12\,
\Lambda\,{r}^{2}-6 =0,\nonumber\\&& \,F_0:= {\it f_1}\, {\it y_1}-
{\it g_1}\, {\it x_1},
\end{eqnarray}
which integrates for ``a middle of the road'' $Q(r)$  as
\begin{eqnarray}
Q \left( r \right) =1/2\,\beta\,\kappa\,{r}^{2} F_0 \left(
\beta\,{r}^{2}-2 \right) +1 /3\,\Lambda\,{r}^{2} \left(
{a}^{2}+{r}^{2} \right) +{r}^{2}-2\,m\,\,r+{C_0},
\end{eqnarray}
where $m$ and $C_0$ are constant of integration. Substituting it
into the second $W$ equation
\begin{eqnarray} &&
W(r)=3\,r^2\, {\frac {d^{2}Q}{d{r}^{2}}}
  -12\, r\,{\frac {d}{dr}}Q  +12\,Q  \nonumber\\&&
-6\,\kappa\, F_0
 \left(1+ \beta\,{r}^{2} \right) +6\,{r}^{2}
+2 \,\Lambda\,{a}^{2}{r}^{2} -12\, a^2  =0,
\end{eqnarray}
one determines the integration constant $C_0$
\begin{eqnarray}
C_0=\kappa\, F_0 /2+{a}^{2}.
\end{eqnarray}
Hence, the structural function $Q(r)$ can be given as
\begin{eqnarray}
Q(r)=\frac{\kappa\,F_0}{2}\,
 \left( \beta\,{r}^{2}-1 \right) ^{2} F_0
 +\frac{\Lambda\,}{3}\,{r}^{2} \left( {a}^{2}+{r}^{2} \right)
+{r}^{2}-2\,m\,r+{a}^{2},
\end{eqnarray}
or, in the representation of $Q(r)$ in terms of the auxiliary
function $K(r)$ from (\ref{solQbeta})
\begin{eqnarray}
K(r)=\frac{\kappa \,F_0}{2}(1-\beta\,r^2)^2 ,
 \end{eqnarray}
in the case one were integrating the Einstein equations for the
function $K(r)$ and its derivatives.

This structure, in certain sense, represents the linear
superposition in $Q$ of different contributions to the matter
tensor; for vacuum, the mass $m$, and the rotation parameter $a$,
$\Lambda$ for  the cosmological constant, the parameters $e$  and
$g_0$ for the electric and magnetic charges through the constant
$F_0$, and finally the $\beta$--parameter responding to the
nonlinearity of the electrodynamics.In general electrodynamics
$L(F,G)$, beside the constant $F_0=f_1y_1-g_1x_1\neq 0$ to have
$L_F\neq 0$, there is a second field constant $G_0:=f_1x_1+g_1y_1$
which takes care of the presence of the second invariant $G $
through the non vanishing of $L_G$ even in the case of the linear
Maxwell field.
 Without any loss of generality, one may set $ {\it f_1}={\it
g_0},{\it g_1}=e$ and equate
\begin{eqnarray}&&
{\it x_1}=-{\frac {{\it F_0}\,e-{\it G_0}\,{\it g_0}} {{e}^{2}+{{\it
g_0}}^{2}}},{\it y_1}={\frac {{\it F_0}\,{\it g_0}+{\it G_0}
\,e}{{e}^{2}+{{\it g_0}}^{2}}}.
\end{eqnarray}
In the Maxwell case, the Kerr--Newman solution is determined for
 $${\it f_1}={\it g_0},{\it g_1}=e,{\it y_1}={\it g_0}, {\it x_1}=-e,
 {\it G_0}=0,{\it F_0}={e}^{2}+{{\it g_0}}^{2},
 $$
with $L(F)= L(F_0)\neq 0$ and $L_G=0$. Therefore, as a by product,
we got the Kerr--Newman solution for a Lagrangian depending on the
two invariants $F\simeq {\bf E}^2-{\bf B}^2$ and $ G\simeq{\bf
E}\cdot{\bf B} $, although it can be determined via duality
rotations \cite{SalazarGarciaPleb1}. \\

It should be pointed out that the equation for the curvature
 scalar
 \begin{eqnarray}&&
 \frac{1}{2}R=OE_{11}-OE_{44} =\kappa[2\,L -\frac{\partial A_t }{\partial \theta  }
 \frac{\partial {}^{\star} P_t }{\partial r  } -\frac{\partial A_t }{\partial r  }
 \frac{\partial {}^{\star} P_t }{\partial \theta  }] +2\Lambda = \Lambda
 +\frac{K^{\prime\prime}}{2\,\rho^2} ,
   \end{eqnarray}
  can be used, as a short cut, to derive the Lagrangian
  $L$, once the vector potential components have been determined, i.e.,
  the components $A_t$ and  ${}^{\star}P_t$ that satisfy the KEY equation.

\section{Curvature quantities $\Psi_2$, $S$, $R$}
The evaluation of the curvature quantities yields:\\
The single Weyl invariant eigenvalue $\Psi_2$, in there dependence
in $Q(r)$, becomes
 \begin{eqnarray}&&
 12\, \left( a\,\cos {\theta } +ir
 \right)  \left( ir-a\,\cos {\theta } a \right) ^{3}\,\Psi_{2Q}
 \nonumber\\&& =-2\,{a}^{2} \left( {a}^{2}\, \cos^{2}
 {\theta} +6\, {r}^{2}\, \cos^{2} {\theta
 }-{r}^{2
}  +4\,i\,a\,r\,\cos {\theta}  \right) \Lambda/3\nonumber\\&&+\,
\left( a\,\cos { \theta }-ir \right) ^{2}{\frac {d^{2}}{d{r}^{2}}}Q
\,+6\, \left( \,i\,a\cos {\theta }+\,r
 \right) {\frac {d}{dr}}Q   -12\,Q
  \nonumber\\&&-8\,i\,a\,r\cos{\theta} -2\,{a}^{2} \cos^{2}
 {\theta} +12\,{a}^{2}+2\,{r}^{2},
\end{eqnarray}
 while in its dependence in the function $K(r)$ simplifies
 considerable
\begin{eqnarray}&&
 12\, \left( a\,\cos{\theta}+ir
 \right)  \left( ir-a\,\cos{\theta} \right) ^{3}
 \Psi_{2K}\nonumber\\&&=  12\,\left( -i\,a\,\cos {\theta }
+r \right) m+\, \left( a\,\cos { \theta }-ir \right) ^{2}{\frac
{d^{2}}{d{r}^{2}}}K \,+ 6\,\left( i\cos { \theta }a+r
 \right) {\frac {d}{dr}}K   -12\,K .
  \end{eqnarray}
For the cubic solution, this $\Psi_2$ quantity becomes
\begin{eqnarray}&&
-12\,{\left(i\,r- a\,\cos {\theta}  \right) ^{3} \left( i\,r+a\cos
\theta \right) }\,\Psi_2(\beta)= 6\,\kappa\,{\it
F_0}\nonumber\\&&-12\,m \left( r -i\,a\,\cos \theta \right)
-6\,{a}^{2} \kappa\,{\it F_0}\,{ \beta}^{2}\,{r}^{2} \cos^{2}
{\theta} \nonumber\\&&+ \left( 2 \,{a}^{2} \cos^{2} {\theta}
-2\,{r}^{2}+8\,i\,a\,r\,\cos {\theta}
 \right) \kappa\,{\it F_0}\,\beta.
\end{eqnarray}
The traceless Ricci tensor eigenvalue $S={S^1}_1=2\,\Phi_{(1\,1)}$,
 in terms of the auxiliary function $K$, amounts to
\begin{eqnarray} &&
{S^1}_1=\,\frac{1}{2\rho^4} \left({\rho}^{2}{\frac
{d^{2}}{d{r}^{2}}}K -4\,r\,{\frac {d}{dr}}K
 +4\,K  \right) ,
\end{eqnarray}
 which for the solution under consideration results in
\begin{eqnarray}\label{Riccibeta}
2\,\rho^4\,S({\beta})&&=\,\kappa\,{\it F_0}\,\beta\, \left(
3\,{a}^{2}\,\beta\,{r}^{2}\,
 \cos^{2} {\theta} -{a}^{2}
  \cos^{2} {\theta}  +{r}^{2} \right) +\,\kappa\,{\it F_0}.
\end{eqnarray}
Finally, the scalar curvature is given by
\begin{equation}\label{scalarCb}
R({\beta})== -2\frac{\Lambda\,a^2}{3\rho^2}
 (6\cos^2{\theta}-1)-\frac{Q^{\prime\prime}-2}{\rho^2} =-4\,\Lambda
-    \frac{K^{\prime\prime}}{\rho^2} =\,2\,{\frac {\kappa\,{\it
F_0}\,\beta\, \left(1- 3\,\beta\,{r} ^{2} \right) }{{a}^{2}  \cos^{
2} {\theta} +{r}^{2}}}-4\Lambda.
\end{equation}

\section{Energy Conditions}
 Remarkable is the simplicity and the invariant character of the
energy conditions which hold for any non linear electrodynamics
energy--momentum tensor of the studied class referred to the
eigenvector orthonormal tetrad frame;
\begin{eqnarray}&&
\mu+p_{\theta}=\frac{2}{\kappa}\,S\geq 0,
\mu-p_{\theta}=\frac{1}{2\kappa}\,R=-\frac{1}{2}\,{T^a}_a\geq
0,\nonumber\\&& \mu+p_{r}=0,\,
 \mu=T_{\mu\nu}u^{\mu}u^{\nu}\geq 0.
\end{eqnarray}
On the other hand, the local energy flow vector $q^a=
  T^{ab}u_b=[0,0,0,-\kappa\,\mu(\theta,r)]
  $ is always a timelike vector except when the energy density vanishes,
  since the norm of the flow vector is
  $q^a\,q_a=-\kappa^2 \mu(\theta,r)^2$.

For our solution these quantities require  the scalar curvature
(\ref{scalarCb}) ${R}(\beta)\geq 0$, and as well
(\ref{Riccibeta}), $S(\beta)={S^1}_1({\beta})\geq 0$.\\
The energy conservation ${T^{\mu\nu}}_{;\nu}=0$ is encoded in the
Einstein equations. The energy density $\mu$ measured by an observer
with 4--velocity, $u^\mu,\,u^\mu\,u_\mu=-1$  is defined as
$\mu(x^\alpha)= T_{\nu\nu}u^\mu\,u^{\nu}$; for the energy--tensor
refereed to the orthonormal tetrad, it occurs to be:
\begin{eqnarray}&&
\kappa\,\mu(\theta,\,r):=OE_{44}=-\frac{\Lambda \,a^2}{3\,\rho^4}
   \left( 3\,a^2\,\cos^4{\theta}+6\,r^2\,\cos^2{\theta}-r^2\,\right)
   \nonumber\\&& -\frac{1}{\rho^4}   \left( r\,Q^{\prime}-Q+a^2-r^2
   \right)=-\Lambda-{\frac {r\,K^{\prime} -K
 }{ \rho^4}}.
\end{eqnarray}
Isolating the derivative $ {\frac {d}{dr}}Q \left( r \right)$,
\begin{eqnarray}&&
 \frac{d Q}{d r}=\frac{1}{r}(Q+r^2-a^2)-\frac{\Lambda a^2}{\,r}
 \left(
 3\,a^2\,\cos^4{\theta}+6\,r^2\,\cos^2{\theta}-r^2\,\right)-\frac{\rho^4}{r}
 \kappa\,\mu(\theta,r),
 \end{eqnarray}
and substituting it into the equation (\ref{OEDD11}), $OE_{11}$, one
arrives at the relation
  \begin{eqnarray}&&
 OE_{11}=-\frac{\rho^2}{2\,r} \frac{\partial}{\partial r}
  \kappa\,\mu(\theta,r)-\kappa\,\mu(\theta,r)
 \end{eqnarray}
  therefore, identifying $OE_{11}=:\kappa\,p_{\theta}$, one has
 \begin{eqnarray}&&
2S= OE_{11}+OE_{44}=\kappa\,p_\theta \left( \theta,r
 \right)+  \kappa\,\mu \left( \theta,r
 \right) =
 -\kappa\,\,\frac {\rho^2 }{2\,r } \frac {\partial}{\partial r}\,\mu \left( \theta,r
 \right)\geq 0
  \end{eqnarray}
 On the other hand from the trace of the Einstein tensor one gets
\begin{eqnarray}&&
 E^\mu_\nu ={ OE^a}_a = { OE^1}_1+{ OE^2}_2+{ OE^3}_3+{
 OE^4}_4=2 { OE}_{11}- 2{ OE}_{44}=-R=\kappa T+4\,\Lambda ,
 \end{eqnarray}
 where it has been taking into account $OE_{22}=-OE_{44}$, and  $OE_{33}=OE_{11}$.
 Thus by the definitions
   $OE_{11}=:\kappa\,p_{\theta}$ and  $OE_{44}=:\kappa\,\mu$  on arrives at
 \begin{eqnarray}&&
\kappa\,\left( \mu \left( \theta,r
 \right) -p_\theta \left( \theta,r
 \right)\right)=\frac{1}{2\,} R.
  \end{eqnarray}
For the solution under consideration one gets:\\
the energy density $\mu(\theta,r)$,
 \begin{eqnarray}&&
 \kappa\,\mu(\theta,r)=-\Lambda-\,{\frac {{\it
F_0}\,\kappa\, \left( 3 \,\beta\,{r}^{2}+1 \right)  \left(
\beta\,{r}^{2}-1 \right) }{2\rho^4}} \geq 0,
 \end{eqnarray}
the scalar curvature $R$,
 \begin{eqnarray}&&{R }=-4\,\Lambda-2\,\frac {\kappa\,{\it F_0}\,\beta\, \left(
3\,\beta\,{r}^{2}-1 \right) }{\rho^{2}}   \geq 0 ,
\end{eqnarray}
the traceless Ricci tensor eigenvalue $S$,
     \begin{eqnarray}
{\it S^1_1} \left( \theta,r \right) =\frac{\kappa\,{\it F_0}\,
}{\rho^4}\left( 3 \,  \cos^{2}{\theta}{a}^{2}{\beta}^{2}{r }^{2}-
\cos^{2} {\theta }{a}^{2}\beta+{r} ^{2}\beta+1 \right) \geq 0 ,
\end{eqnarray}
finally the pressure $ p_\theta\left( \theta,r \right)\geq 0$,
  \begin{eqnarray}
\kappa\, p_\theta\left( \theta,r
\right)=OE_{11}=\frac{1}{2\rho^4}\,\kappa\,{\it F_0}\,
 \left( 6\,{a}^{2}
\cos^{2} {\theta }
\,{\beta}^{2}{r}^{2}+3\,{\beta}^{2}{r}^{4}-2\,{a}^{2}  \cos^{2}
 {\theta} \beta+1 \right) .
\end{eqnarray}

 \section{Final remarks}

 In this work we presented in detail the derivation of the stationary
 axisymmetric black hole solution to the Einstein equations coupled to
 nonlinear electrodynamics in the presence of a cosmological constant
 of both signs. This solution possesses mass $m$, rotation parameter $a$,
 electric and magnetic charges $e$ and $g_0$, nonlinear electrodynamics
 parameter $\beta$, and two parameters $F_0$ and $G_0$ associated to
 the presence of the invariants $F$  and $G$ in the Lagrangian $L$,
 and finally a cosmological constant $\Lambda$ for de Sitter and Anti de
 Sitter branches of solution. In the forthcoming works of the series
 we deal mostly with the physical interpretation of the solution:
 the study the trajectories of neutral and charged test particles,
 the trajectories of light rays, the birefringence of light, the
 black hole properties of the solution; horizons, maximal
 extension, Penrose diagrams, and thermodynamics, among others.

\end{document}